**Title: Precision Medicine for the Population—The Hope and Hype of Public Health Genomics**

**Authors:**

JunBo Wu, BA,[1] Nathaniel C. Comfort, PhD[2]

1. Department of History of Science and Technology, Zanvyl Krieger School of Arts and Sciences, Johns Hopkins University, Baltimore, Maryland
2. Department of History of Medicine, Johns Hopkins University School of Medicine, Baltimore, Maryland

**Corresponding Author:**

JunBo Wu, BA

Department of History of Science and Technology

Zanvyl Krieger School of Arts and Sciences

Johns Hopkins University

3400 N. Charles Street, Gilman Hall 301, Baltimore, MD 21218

Email: jwu149@jh.edu

**Abstract:**

Public health is the most recent of the biomedical sciences to be seduced by the trendy moniker "precision." Advocates for "precision public health" (PPH) call for a data-driven, computational approach to public health, leveraging swaths of genomic "big data" to inform public health decision-making. Yet, like precision medicine, PPH risks overselling the value of genomic data to determine health outcomes, but on a population-level. A large historical literature has shown that over-emphasizing heredity tends to disproportionately harm underserved minorities and disadvantaged communities. By comparing and contrasting PPH with an earlier attempt at using big data and genetics, in the Progressive era (1890–1920), we highlight some potential risks of a genotype-driven preventive public health. We suggest that such risks may be avoided by prioritizing data integration across many levels of analysis, from the molecular to the social.

Precision medicine has hit the target but missed the bullseye. Since before the Human Genome Project, scientists have promised that our DNA "instruction book" will tailor medicine to our unique physiologies, refashioning "one-size-fits-all" care that has prevailed since the nineteenth century.[1] Genomics has led to improved treatments for a number of rare diseases. Yet application to common diseases, which tend to have complex causes involving many interacting genetic and environmental inputs has proven difficult, prompting critics to question whether precision medicine can live up to its hype.[2,3] Despite an increasingly precise read of our DNA, efforts to translate that knowledge into treatment are all too often wide of the mark.

**Commented [NC1]:** This could use a footnote to a review article in a respected journal

Meanwhile, "precision" approaches have multiplied across the biomedical sciences. Precision oncology, nutrition, and agriculture are just a few fields hoping to be revolutionized by genomics.[4-6] Jumping on the bandwagon, one company even offered precision services to wine-tasting based on one's genome profile.[7]

In some ways, one of the strangest applications of the "precision" moniker is to public health. In 2016, Muin Khoury, Director of the CDC's Office of Genomics, first called for a "precision public health" (PPH).[8,9] Since then, Khoury, the Gates Foundation, and other public health leaders have pushed their chips into the genomics pot, upping the ante for public health informed by genomic data. Their bet is on transforming population health management with genomics, big data, and machine learning.[10] For example, in 2020, researchers in the Netherlands combined whole genome-sequence analysis with epidemiological data to model and predict SARS-CoV-2 community transmission patterns, improving contact-tracing efficiency.[11] For Khoury, the PPH end goal is a world that unites public health policy with data-intensive analytics.[12]

Yet, public health is the poster child for one-size-fits-all medicine. A product of the germ theory of disease launched by Louis Pasteur and Robert Koch in the mid-nineteenth century, the discovery of specific disease agents (germs) led to the dream of universal therapy for a given disease. It also represented the full-throated embrace of science by medicine and, not incidentally, produced the biggest therapeutic revolution in the history of medicine.[13] "One size fits all" is nothing to be ashamed of!

No branch of medicine benefited more from the germ theory than public health. Flu shots, mosquito nets, and sanitation are low-cost solutions that work for (almost) everyone. Not for nothing, then, has public health identified with slogans like, "Protecting Health, Saving Lives—*Millions at a Time*."[14] d biobanks into public health necessarily lead to a "precision" public health?[15] Or, in embracing the fad of precisionism, is public health almost literally throwing the baby out with the foul, disease-causing bathwater?

These inherent contradictions have pushed PPH to upsell the added value genomics can bring. Major chronic diseases, including cardiovascular disease and obesity, alongside breakthroughs in understanding genetic variation in infection disease susceptibility, have opened opportunities for using human genomics to better combat major public health crises of the 21st century.[16] Most recently, Khoury argued how initiatives to understand our predisposition to COVID-19 demonstrates the unique value of integrating genetic risks. [17] A tech-centric public health, PPH emphasizes the "nature"-side of nature-nurture.

The dream of using human genetics to improve public health is not new. At least since the Progressive Era (1890–1920), physicians, public-health leaders, and scientists have imagined treating and preventing otherwise incurable hereditary diseases at a population-level. For example, in the 1910s, the Yale economist Irving Fisher proposed a softer sense of heredity as a general complement to the environment. Grafting the contemporary craze for hygiene, newly grounded in the germ theory of disease, onto the simultaneous craze for eugenics, Fisher hoped to prevent disease at all levels, from the individual to the population to the "race" (read alternately as "American," "white," or "human")—that is, the "hygiene of future generations."[18] Fisher was clear that "race hygiene" was a synonym for eugenics.[18] He was the lead author behind the widely read *Report on the National Vitality* (1909), which estimated that improving human "heredity and hygiene" would save the Federal Treasury more than $1.5 billion dollars—more than $45 billion today.[19]

To distinguish population-level patterns of heredity, eugenicists sought unprecedented amounts of data—a sort-of "big data" for their times. In 1910, Harvard-trained biologist Charles Davenport established the Eugenics Record Office (ERO), with esteemed support from the likes of Alexander Bell and the Carnegie Institution of Washington.[20] Emboldened by the ERO's research and propaganda, public health experts, psychiatrists, and criminologists systematically rounded-up and sterilized tens of thousands of "feebleminded" (adults with a mental age below 12), syphilitic, and epileptic Americans, on grounds of "preventative medicine" for the supposed illness.[21] Pioneering the public-health surveillance of heredity, the ERO hired hundreds of mostly young women as "fieldworkers" to collect genetic data on the diseased and insane, much as public health officials canvassed low-income neighborhoods for unhygienic, disease-spreading households. [22] By the mid-1920s, over 750,000 of these eugenic records were completed.[23] Like Davenport, Raymond Pearl of the public health school at Johns Hopkins assembled pedigrees and records through fieldworkers, adding more requirements for clinical measurements and examinations to genetic factors of complex diseases, such as tuberculosis and hypertension.[24] The datasets grew in both size and depth, with more information spanning a patient's biological and clinical backgrounds. A century ago, as today, it made sense to begin the data analysis with genes.

More data made it easier for eugenicists to find pedigree patterns to support the idea of the genetic supremacy of the privileged classes. The explosive rise of Mendelian genetics in this period spurred Davenport, Pearl, and others to consider genetic explanations—and even single-gene causes—for complex mental illnesses and diseases. As genetics offered ever simpler, more reductionist explanations for disease, public health and preventive medicine became increasingly deterministic about innate, hereditary predisposition, downplaying the role of environmental factors. For example, Charles V. Chapin, the health commissioner of Rhode Island, abruptly reversed his sanitationist stance for one that blamed public-health problems on individuals'—especially immigrants'—faulty germ plasm. [24,25]

In short, history shows that big data is not a panacea against over-reductionist interpretations and loss of attentiveness to the individual. It may even encourage a misleading genetic determinism. Progressive-era race hygiene was hampered more than helped by its over-emphasis on heredity, its neglect of what we now call the social determinants of health. For Fisher, heredity and hygiene embraced, even integrated, both nature and nurture. But with the triumph of Mendelism in the first decades of the 20th century, the holistic approach lost out to reductionist explanations of complex traits, which explained even intelligence in terms of a simple recessive "gene for." And this shift occurred even as genetic datasets swelled to unprecedented size. Finally, the eugenicists' biased interpretations were more likely to match ungeneralizable patterns from a growing dataset, reinforcing their determination to substantiate genetic determinism.

As with a century ago, reductionism blurs into determinism if we choose, intentionally or not, to neglect the social determinants of health. What's more, as our individuality and experiences are reduced to quantifiable bits and bytes in the information age, a *data* determinism may prevail with increased reliance on algorithmic predictions. As scholars have shown, our most advanced AI are only as good as the data we feed them.[26,27] Without care, data-driven tools can perpetuate structural inequities ingrained in the data.

How could the social determinants of health be folded into a "precision" public health? By bringing "big data" thinking to a broad range of quantifiable social and environmental risk factors. For example, high-spatial resolution maps have revealed otherwise unknown neighborhood-level disparities in child mortality within 46 different African countries, assisting resource allocation to communities most in need.[28] At NYU recently, researchers summarized a key list of data sources and algorithms with most potential to capture the physical, economic, and environmental conditions relevant to public health practice.[29] Such efforts could even enhance the value of genomic data by facilitating the investigation of gene–environment interactions—the biggest share of causes of complex traits. Such a multi-level approach further requires representative team members who are as informed in data analytics as they are with ethical challenges in public health.

Instead of looking at what genomics can do individually, then, PPH should assess how genomic data can enhance population-level risk factors. In the future of public health, genomics will make the ways in which we understand population risks more precise. To create effective interventions and avoid determinism, however, transdisciplinary research grasping the entire context is needed to make precision public health more accurate.

**Acknowledgements**

The authors have no conflicts of interest.

**References**

1. Collins FS. Medical and Societal Consequences of the Human Genome Project. *The New England Journal of Medicine.* 1999;341(1):28-37. doi:10.1056/NEJM199907013410106

2. Comfort N. The Overhyping of Precision Medicine. www.theatlantic.com Web site. https://www.theatlantic.com/health/archive/2016/12/the-peril-of-overhyping-precision-medicine/510326/. Updated 2016. Accessed Nov 16, 2022

3. Aronowitz R. What Is Medicine For? www.bostonreview.net Web site. https://www.bostonreview.net/articles/robert-aronowitz-what-medicine/. Updated 2019. Accessed Nov 17, 2022

4. Prasad V, MPH, Fojo T, Prof, Brada M, Prof. Precision oncology: origins, optimism, and potential. *The lancet oncology.* 2016;17(2):e81-e86. doi:10.1016/S1470-2045(15)00620-8

5. Heianza Y, Qi L. Gene-Diet Interaction and Precision Nutrition in Obesity. *International journal of molecular sciences.* 2017;18(4):787. doi:10.3390/ijms18040787

6. Monteiro A, Santos S, Gonçalves P. Precision Agriculture for Crop and Livestock Farming—Brief Review. *Animals (Basel).* 2021;11(8):2345. doi:10.3390/ani11082345

7. Robbins R. Fruity with a hint of double helix: A startup claims to tailor wine to your DNA. STAT, Oct 27, 2016. https://search.proquest.com/docview/2618184234

8. Khoury MJ. Precision Public Health: What Is It? blogs.cdc.gov Web site. https://blogs.cdc.gov/genomics/2018/05/15/precision-public-health-2/. Updated 2018. Accessed Nov 17, 2022

9. Khoury MJ. Precision Public Health and Precision Medicine: Two Peas in a Pod. blogs.cdc.gov Web site. https://blogs.cdc.gov/genomics/2015/03/02/precision-public/. Updated 2015. Accessed Nov 17, 2022

10. Khoury MJ, Holt KE. The impact of genomics on precision public health: beyond the pandemic. *Genome medicine.* 2021;13(1):67. doi:10.1186/s13073-021-00886-y

11. Oude Munnink BB, Nieuwenhuijse DF, Stein M, et al. Rapid SARS-CoV-2 whole-genome sequencing and analysis for informed public health decision-making in the Netherlands. *Nature medicine.* 2020;26(9):1405-1410. doi:10.1038/s41591-020-0997-y

12. Khoury M, Engelgau M, Chambers D, Mensah G. Beyond Public Health Genomics. *Public health genomics.* 2018;21(5/6):244-249. doi:10.1159/000501465

13. Vogel MJ, Rosenberg CE. *The Therapeutic Revolution.* University of Pennsylvania Press; 1979

14. Johns Hopkins Bloomberg School of Public Health. Mission, Vision, and Values. publichealth.jhu.edu Web site. https://publichealth.jhu.edu/about/at-a-glance/mission-vision-and-values. Accessed Nov 19, 2022

15. Chowkwanyun M, Bayer R, Galea S. “Precision” Public Health — Between Novelty and Hype. *The New England journal of medicine.* 2018;379(15):1398-1400. doi:10.1056/NEJMp1806634

16. Roberts MC, Fohner AE, Landry L, et al. Advancing precision public health using human genomics: examples from the field and future research opportunities. *Genome medicine.* 2021;13(1):97-107. doi:10.1186/s13073-021-00911-0

17. Khoury MJ, Gwinn M, Duggal P. The Public Health Impact of COVID-19: Why Host Genomics? 2020

18. Fisher I, Fisk EL, Life Extension Institute I. *How to Live: Rules for Healthful Living, Based on Modern Science.* Funk and Wagnalls; 1915

19. Fisher I. *Bulletin 30 of the Committee of One Hundred on National Health: Being a Report on National Vitality, Its Wastes, and Conservation.* U.S. Government Printing Office; 1909

20. Greenwald BH. *Alexander Graham Bell through the Lens of Eugenics, 1883—1922.* George Washington University; 2006

21. MacDowell EC. Charles Benedict Davenport, 1866-1944: A Study of Conflicting Influences. *Bios.* 1946;17(1):3-50. https://www.jstor.org/stable/4604920

22. Petty TL. The history of COPD. *International Journal of Chronic Obstructive Pulmonary Disease.* 2006;1(1):3-14. doi:10.2147/copd.2006.1.1.3

23. Bix AS. Experiences and Voices of Eugenics Field-Workers: 'Women's Work' in Biology. *Social Studies of Science.* 1997;27(4):625-68. doi:10.1177/030631297027004003

24. Comfort N. *The Science of Human Perfection: How Genes Became the Heart of American Medicine.* Illustrated ed. Yale University Press; 2014

25. Pernick MS. Contagion and Culture. *American literary history.* 2002;14(4):858-865. doi:10.1093/alh/14.4.858

26. Noble SU. *Algorithms of oppression: how search engines reinforce racism.* New York University Press; 2018. https://www.jstor.org/stable/j.ctt1pwt9w5

27. Benjamin R. *Race after technology: abolitionist tools for the new Jim code.* Polity; 2019

28. Golding N, Burstein R, Longbottom J, et al. Mapping under-5 and neonatal mortality in Africa, 2000–15: a baseline analysis for the Sustainable Development Goals. *Lancet (London, England).* 2017;390(10108):2171-2182. doi:10.1016/S0140-6736(17)31758-0

29. Thorpe L, Chunara R, Roberts T, et al. Building Public Health Surveillance 3.0: Emerging Timely Measures of Physical, Economic, and Social Environmental Conditions Affecting Health. *American journal of public health (1971).* 2022;112(10):1436-1445. doi:10.2105/AJPH.2022.306917